\documentclass[prl,twocolumn,showpacs,preprintnumbers,amsmath,amssymb,superscriptaddress]{revtex4}
\date{\today}

\usepackage{graphicx} 
\usepackage{dcolumn}  
\usepackage{bm}       

\begin{document}

\title{Fission Barriers of Compound Superheavy Nuclei}
\author{J.C. Pei}
\affiliation{Joint Institute for Heavy Ion Research, Oak Ridge, TN
37831}
 \affiliation{Department of Physics and
Astronomy, University of Tennessee Knoxville, TN 37996}
\affiliation{Physics Division, Oak Ridge National Laboratory, P.O.
Box 2008, Oak Ridge, TN 37831}
\author{W. Nazarewicz}
 \affiliation{Department of Physics and
Astronomy, University of Tennessee Knoxville, TN 37996}
\affiliation{Physics Division, Oak Ridge National Laboratory, P.O.
Box 2008, Oak Ridge, TN 37831}
\affiliation{Institute of
Theoretical Physics, Warsaw University, ul. Ho\.{z}a 69, PL-00681
Warsaw, Poland}
\author{J.A. Sheikh }
\affiliation{Department of Physics and Astronomy, University of
Tennessee Knoxville, TN 37996}
\affiliation{Physics Division,
Oak Ridge National Laboratory, P.O. Box 2008, Oak Ridge, TN 37831}
\author{A.K. Kerman}
\affiliation{Department of Physics and Astronomy, University of
Tennessee Knoxville, TN 37996}
\affiliation{Physics Division,
Oak Ridge National Laboratory, P.O. Box 2008, Oak Ridge, TN 37831}
\affiliation{Center for Theoretical Physics,
Massachusetts Institute of Technology,
Cambridge, MA 02139}

\begin{abstract}
The dependence of fission barriers on the excitation energy of the
compound nucleus impacts the survival probability of superheavy nuclei
synthesized in heavy-ion fusion reactions. In this work, we investigate
the isentropic fission barriers by means of the self-consistent nuclear
density functional theory. The relationship between isothermal and
isentropic descriptions is demonstrated. Calculations have been carried
out for $^{264}$Fm, $^{272}$Ds, $^{278}$112, $^{292}$114, and
$^{312}$124. For nuclei around $^{278}$112 produced in ``cold fusion"
reactions, we predict a more rapid decrease of fission barriers with
excitation energy as compared to the nuclei around $^{292}$114
synthesized in ``hot fusion" experiments. This is explained in terms of
the difference between the  ground-state and saddle-point temperatures.
The effect of the particle gas is found
to be negligible in the range of temperatures studied.
\end{abstract}


\maketitle

What are the heaviest nuclei that can exist? To answer this question,
nuclear physicists explore superheavy systems
at the limit of mass and charge. During recent years,
the field
has witnessed remarkable progress \cite{[Hof00],[Oga07],[Lov07]}
in the production and identification of new elements.
The major experimental challenge is to find optimal
beam-target combinations and kinematic conditions  that would   lead to the formation,
at reasonable rates, of the species of interest.
One of the key problems is the survival probability
of a  superheavy nucleus synthesized in a heavy-ion fusion reaction that
depends on a competition between fission and
particle evaporation \cite{[Swi08]}.

The dependence of the  fission
barrier on the excitation energy  is  among the key factors determining
the production of a superheavy  nucleus.
Since shell effects, essential for the mere existence of superheavy nuclei,
are quenched at high temperatures
(see, e.g., Refs.~\cite{[Ign80],[Egi00]}),
it is expected that the fission barriers in superheavy CN should
quickly decrease with  energy. In the  analysis of experimental data, this is usually accounted for by a
phenomenological damping factor \cite{[Arm00a],[Den00],[Itk02],[Lov07]}
(cf. discussion in Ref.~\cite{[Swi08]}).

Microscopically,  shell effects in superconducting heated nuclei can
be self-consistently treated by the Finite-Temperature
Hartree-Fock-Bogoliubov (FT-HFB) method
\cite{[Egi00],[Mar03],[Kha07],[Min08]}. For superheavy nuclei,
although there have been extensive self-consistent studies of
zero-temperature fission pathways, (see, e.g.,
\cite{[Bur04],[Ben98a],[Sta05]}), studies of CN fission  have been
virtually nonexistent. Moreover, in the majority of studies, CN
fission has been treated as an isothermal process in terms of free
energy, $F$=$E-TS$ at a fixed temperature $T$. The assumption of
$T$=const is certainly wrong:  the fissioning CN is not connected to
an external thermal reservoir. Physically, a more appropriate
picture of fission is that governed by the  isentropic process
\cite{[Die81]}. The necessary condition for the isentropic scenario
is that the collective motion is adiabatic, i.e.,  no heat energy
can be delivered to nor extracted from the system.

The aim of this study is to investigate self-consistent isentropic
fission pathways in superheavy CN, with a focus on energy dependence
of fission barriers. To this end, we selected several  nuclei of
current experimental interest: (i) $^{272}$Ds  and $^{278}$112$^{*}$
that have been synthesized in the ``cold-fusion" reaction using a
$^{208}$Pb target at excitation energies $E^{*}$ of $\sim$10-12 MeV
\cite{[Arm00a]}; (ii) the nucleus $^{292}$114$^{*}$ produced in the
``hot-fusion" reaction $^{48}$Ca+$^{244}$Pu at
$E^{*}$$\sim$36-40\,MeV  \cite{[Oga00a]}; (iii) the nucleus
$^{312}$124$^{*}$ at $E^{*}$$\sim$80\,MeV studied by means of
$^{74}$Ge+$^{238}$U reaction and crystal blocking \cite{[Mor08]};
and (iv) the $^{264}$Fm that is expected to fission symmetrically
into two doubly-magic $^{132}$Sn nuclei. The FT-HFB calculations are
carried out using the two Skyrme-HFB codes: the recently developed
axial coordinate-space solver HFB-AX \cite{[Pei08]} and a
symmetry-unrestricted   solver HFODD \cite{[Dob04c]}. The
description of thermal  properties  involves significant
contributions from high-lying single-particle states which give rise
to the particle gas as the temperature increases \cite{[Bon85c]},
requiring a very large configuration space to guarantee convergence
\cite{[Oko87]}. In this respect, HFB-AX is an excellent tool as it
allows calculations in very large deformed boxes. The
finite-temperature formalism has been implemented in HFB-AX and
HFODD in the usual way \cite{[Kha07]} by introducing the
thermal-averaged particle and pairing densities through the Fermi
distribution function.

In the particle-hole channel, we use the  SkM$^*$ energy density
functional \cite{[Bar82]} that
 has been optimized  at large deformations; hence it is often used
for fission barrier predictions. In the pairing channel, we adopted
 the
density-dependent $\delta$
 interaction in the mixed variant \cite{[Dob02c]}. The pairing strengths,
$V_p$=$-$332.5 MeV\,fm$^{-3}$ (protons) and $V_n$=$-$268.9
MeV\,fm$^{-3}$ (neutrons) have been fitted to reproduce the  pairing
gaps in $^{252}$Fm. The details of HFB-AX calculations follow
Ref.~\cite{[Pei08]}. We used $M$=13 order B-splines, and the maximum
mesh size $h$=0.6 fm.  The cylindrical box employed depends on the
total quadrupole moment of the system, $Q_{20}$. That is, for
$Q_{20}$$\leq$30\,b we used a square box of $R_\rho$=$R_z$=20.4\,fm;
for 30$<$$Q_{20}$$\leq$80\,b we took $R_\rho$=19.2\,fm and
$R_z$=21.6\,fm; and for $Q_{20}$$>$80\,b we took $R_\rho$=18\,fm and
$R_z$=22.8\,fm. The  calculations with HFODD were carried out in a
space of the lowest 1161Ê stretched oscillator  states originating
from the
 31 principal oscillator shells.

\begin{figure}[htb]
 \centerline{\includegraphics[trim=0cm 0cm 0cm
0cm,width=0.32\textwidth,clip]{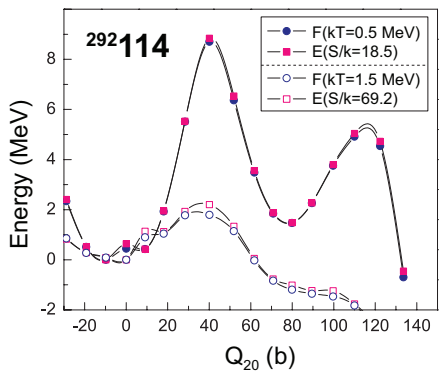}}
 \caption{\label{barriers-a}(Color online) Calculated isothermal
 ($F$ at constant $T$) and
 isentropic ($E$ at constant $S$) axial symmetric
 fission pathways for $^{292}114$
 as a function of the total quadrupole
moment $Q_{20}$. The isothermal calculations
were carried out at
$kT$=0.5 and 1.5\,MeV. In the
isentropic description, $S$ was fixed at the
free energy minimum, i.e., $S/k$=18.5 (69.2) at $kT$=0.5 (1.5)\,MeV.}
\end{figure}
As pointed out in Ref.\,\cite{[Die81]},
isothermal and isentropic
descriptions
can be related  by making use of the thermodynamical identity
$\left(\frac{\partial E}{\partial
Q_{20}}\right)_S=\left(\frac{\partial F}{\partial Q_{20}}\right)_T$
which is reminiscent of the well-known relation
 $(\frac{\partial E}{\partial
V})_S=(\frac{\partial F}{\partial V})_T$.
 Indeed,
$Q_{20}$ enters the variations  $dF$ and $dE$  via the term $-q_{20}
dQ_{20}$, where $q_{20}$ is the  Lagrange multiplier corresponding
to the constraint on the quadrupole moment. Figure~\ref{barriers-a}
displays isothermal  and isentropic axial fission pathways  of
$^{292}$114 as a function of $Q_{20}$.
In the
isentropic description, $S$ was fixed at the value
$S=S(T)$
corresponding to the
free energy minimum at temperature $T$. This was done by performing
constrained FT-HFB calculations for a number of temperatures and
inverting the relation $S=S(T)$ numerically by using  interpolation.
It is seen that the
isothermal and isentropic curves are very close. It is worth noting
that in the macroscopic-microscopic calculations of
Ref.~\cite{[Die81]} the isentropic barriers  are predicted to be
significantly higher than the isothermal ones. The reason for this
is the violation of self-consistency in the macroscopic-macroscopic
theory. Figure~\ref{barriers-a}  shows that the variational
principle behind FT-HFB guarantees practical equivalence of
isothermal and isentropic pictures. The remaining small discrepancy
is due to the numerical interpolation error caused by extraction of
$(E,S)$ values from the original $(F,T)$ mesh.

\begin{figure}[htb]
 \centerline{\includegraphics[trim=0cm 0cm 0cm
0cm,width=0.35\textwidth,clip]{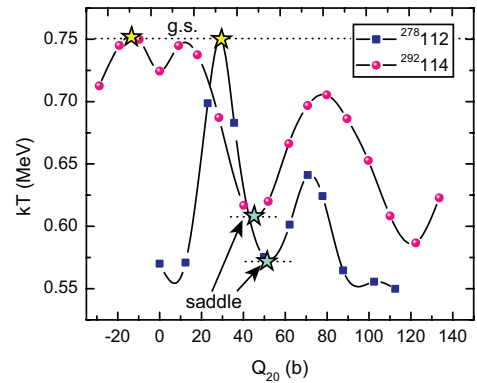}} \caption{\label{temperature-b}
(Color online) The temperature along the isentropic symmetric
fission pathways of
$^{278}$112 and $^{292}$114. The ground state (g.s.) and saddle point
configurations are marked by stars.
The g.s. temperature
was assumed to be
$T_{g.s.}$$\sim$750\,keV  and $S=S(T_{g.s.})$.
The temperature at the saddle point is considerably lower than $T_{g.s.}$.}
\end{figure}
The behavior of temperature $T=T(S)$ is shown in Fig.~\ref{temperature-b}
along the fission pathways of $^{278}$112 and $^{292}$114. The
entropy corresponds to the g.s. value at
$T_{g.s.}$$\sim$750\,keV. It is seen that $T$ changes as a function of $Q_{20}$.
In particular, the barrier temperature is significantly lower than $T_{g.s.}$.
In the following, we shall stick to the
isentropic description, i.e., the
temperature will be related to the  g.s. excitation energy.

\begin{figure}[htb]
 \centerline{\includegraphics[trim=0cm 0cm 0cm
0cm,width=0.40\textwidth,clip]{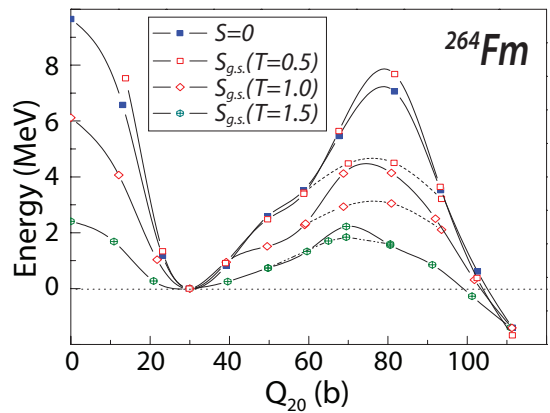}} \caption{\label{barrier-b}
(Color online) Symmetric isentropic fission pathways of $^{264}$Fm at the values of $S$ corresponding to $kT_{g.s.}$=0, 0.5, 1.0, and 1.5\,MeV The  energy
 has been normalized to zero at the ground-state minimum. The
effect of triaxial degrees of freedom on the first barrier is marked by dashed lines.}
\end{figure}
Figure \ref{barrier-b} displays the  energy curves  of $^{264}$Fm as functions
of $Q_{20}$ at different entropies  corresponding to
different values of  $T_{g.s.}$. At $S(kT_{g.s.}=0.5)$ , the
barrier increases by about 0.6\,MeV as compared to $S$=0, due to the
reduction  of pairing correlations
\cite{[Kha07],[Min08],[Egi00],[Mar03]}. (In our calculations,
pairing energies are unimportant above  $kT$=0.7\,MeV.) At higher
excitation energies, the  barrier is gradually reduced
to 0.9 MeV at
$S(kT_{g.s.}=1.5)$, due to the thermal quenching  of shell effects.
In order to estimate the reduction of the fission barrier due to triaxiality
 expected in
the Fm isotopes \cite{[Bon04],[Sta05]}, we performed
symmetry-unrestricted calculations with HFODD. The result is shown
in Fig.~\ref{barrier-b} by a dashed line. At low excitation energies,
triaxiality reduces the fission barrier by $\sim$3-4\,MeV, but the
triaxial shell effect is washed out with increasing entropy and becomes
negligible at  the largest excitations considered.  For the
systematic calculations of triaxial and reflection-asymmetric deformations along the isentropic fission pathways of superheavy nuclei, we refer the reader to
Ref.~\cite{[She08]}.

\begin{figure}[htb]
\centerline{\includegraphics[trim=0cm 0cm 0cm
0cm,width=0.45\textwidth,clip]{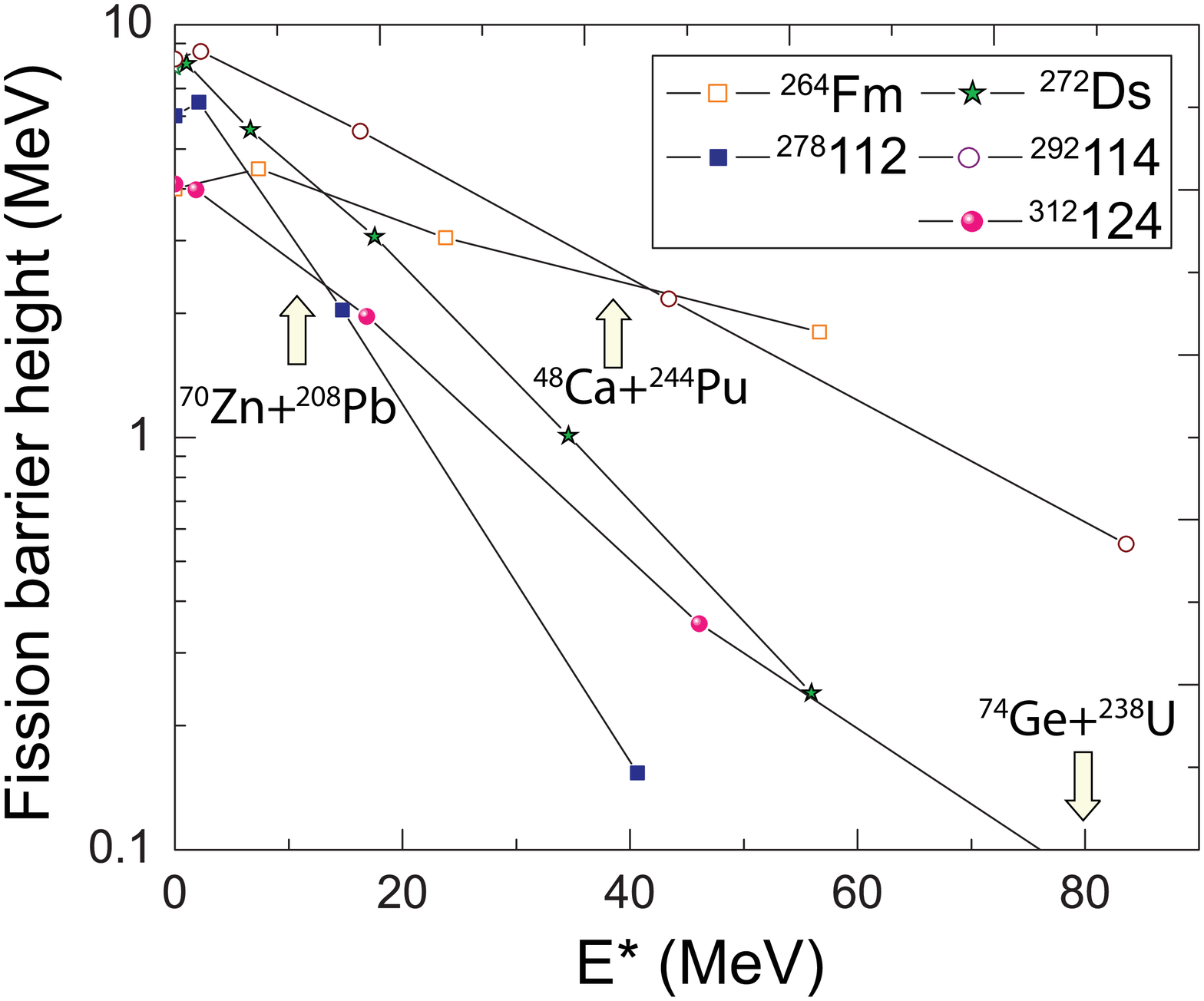}}
\caption{\label{height}
(Color online) The height of the inner fission barrier
in  $^{264}$Fm, $^{272}$Ds, $^{278}$112, $^{292}$114, and
$^{312}$124  as a function of excitation energy $E^*$.
The effect of triaxiality on fission barrier has been included.
The experimental values of $E^*$ corresponding to CN formed in the reactions
indicated are marked by arrows.
}
\end{figure}
To study the excitation energy dependence of fission barriers in more
detail, in Fig.~\ref{height} we plot the  height of the inner
axial fission barrier
of  $^{264}$Fm,  $^{272}$Ds, $^{278}$112, $^{292}$114, and
$^{312}$124 as a function of  the excitation energy
$E^{*}$=$E(S)-E(S=0)$.
 Above $kT_{g.s.}$=0.5\,MeV,
fission barriers $E_B$ are damped exponentially with $E^{*}$:
$E_B$$\propto$$e^{-\gamma_DE^{*}}$.
The value of the damping parameter
$\gamma_D$ is not known well (see discussion in Ref.~\cite{[Den00]})
but $\gamma_D^{-1}$ is usually
taken in the range of 8-20 MeV~\cite{[Den00],[Itk02]}.
According to our calculations, in the CN $^{272}$Ds and $^{278}$112
synthesized in cold fusion reaction
$\gamma_D^{-1}$=17.2 and 10.8\,MeV, respectively, while it is 30\,MeV
in  $^{292}$114 and 20.2\,MeV in $^{312}$124.

The appreciable change in $\gamma_D$  with $N$ and $Z$,
e.g.,
when going from $^{278}$112 to $^{292}$114,
can be traced back
to shell effects. As seen in Fig.~\ref{temperature-b},
in the isentropic picture,
the saddle point  temperature $T_B$ is {\it lower} than
$T_{g.s.}$, i.e., $\Delta T$=$T_{g.s.}-T_B$$>$0 (in a nice analogy to
an adiabatically expanding gas).
 Due to shell effects, $\Delta T$($^{278}$112)$>$$\Delta T$($^{292}$114)
and the  thermodynamical identity
$\left(\frac{\partial E}{\partial
S}\right)_{Q_{20}}=kT$
implies a
larger $\gamma_D$ in $^{278}$112, thus explaining the
pattern seen in Fig.~\ref{height}.

\begin{figure}[htb]
\centerline{\includegraphics[trim=0cm 0cm 0cm 0cm,width=0.40\textwidth,clip]{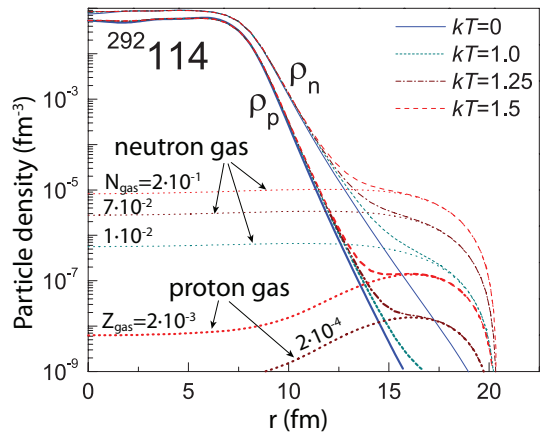}}
\caption{\label{dens}(Color online) Proton (thick lines) and neutron
(thin lines) spherical density
distributions in $^{292}$114 calculated at different temperatures
(in MeV). The
external gas contributions are marked by dotted lines.
The number of nucleons in the gas is indicated by numbers.
The size of the box
in HFB-AX was 20.4\,fm.}
\end{figure}
As discussed in Ref.~\cite{[Ker81]}, the FT-HFB solution in a confined
box (or in a finite localized basis) corresponds to a nucleus located at
the center surrounded by the external gas. The gas produces the pressure
necessary to obtain an equilibrium with the particle-decaying hot
nucleus; the corresponding particle-decay lifetime is in fact inversely
proportional to the  density of the external gas \cite{[Bon85c]}.
Figure~\ref{dens} illustrates the effect of the gas for spherical
configuration in $^{292}$114 at different temperatures. In order to
separate gas contributions, we applied the procedure of
Refs.~\cite{[Bon85c],[Kru00b]}. It is seen that with increasing
temperature, the  gas gradually appears. As expected, the neutron gas is
uniformly distributed within the volume of the box while the proton gas
appears outside the nuclear surface due to the
effect of the Coulomb barrier. The contribution from the gas to the
kinetic energy is very small: it is about 0.2 (0.9)\,MeV at $kT$=1.25 (1.5)\,MeV.
The number of gas neutrons is very small; it
increases from $N_{gas}$=0.01 at $kT$=1\,MeV
to 0.21 at $kT$=1.5\,MeV, cf. Fig.~\ref{dens}.
Consequently, the gas contribution to the
deformation energy is practically negligible in the range of
temperatures considered. At very high temperatures, however, the gas is
expected to significantly change  the Coulomb energy, the total entropy,
and the chemical potentials, and its contribution should be properly
removed~\cite{[Bon85c]}.

In conclusion, we performed self-consistent calculations of
isentropic fission barriers of compound superheavy nuclei based on a
coordinate-space FT-HFB method. We first demonstrated the
relationship between  the isothermal  and isentropic description of
fission and emphasized the role of self-consistency. The conclusion,
important for practical applications, is that the surfaces of
$F(T=T_0)$ and $E(S=S_0)$  in the space of collective coordinates
are identical for $S_0=S(T_0)$ if the  self-consistency condition is
met. It is to be noted,  however, that  this formal connection does
not indicate that the isothermal and isentropic pictures of fission
are similar. The isothermal approach to fission is clearly incorrect
as the nucleus is not connected to a thermal bath, i.e., the
temperatures of the g.s. configuration and the barrier obviously
differ. On the other hand, if the fission process is adiabatic, the
isentropic description should be closer to reality.

Secondly, we demonstrate that the  dependence of isentropic fission
barriers on excitation energy changes rapidly with particle number,
pointing at the importance of shell effects even at large excitation
energies characteristic of compound nuclei. For instance, fission
barriers for $^{272}$Ds and $^{278}$112, produced in a cold fusion
reaction, and $^{292}$114, synthesized in a hot fusion reaction, are
predicted to exhibit markedly different behavior. For  the CN
$^{312}$124, we calculate no isentropic fission barrier at
$E^{*}$$\sim$80\,MeV.

Finally, we show that the external particle gas has no effect on the
fission barriers up to at least $kT$=1.5\,MeV. The barrier damping
parameters extracted from our FT-HFB calculations, as well as the
neutron decay rates extracted from the magnitude of the neutron gas
component \cite{[Bon85c]} can be used  to provide reliable theoretical
estimates of CN survival probability. Work along these lines is in
progress.

Useful discussions with W. Loveland, J. Skalski,  and A. Staszczak are
gratefully acknowledged. This work was supported in part by the National
Nuclear Security Administration under the Stewardship Science Academic
Alliances program through  Grant DE-FG03-03NA00083; by the U.S.
Department of Energy under Contract Nos. DE-FG02-96ER40963 (University
of Tennessee), and DE-AC05-00OR22725 with UT-Battelle, LLC (Oak Ridge
National Laboratory), and DE-FC02-07ER41457 (UNEDF SciDAC
Collaboration). Computational resources were provided by the National
Center for Computational Sciences at Oak Ridge National Laboratory.


\end{document}